\documentclass[twocolumn,showpacs,amsmath,amssymb,aps,prl,floats,nofootinbib,superscriptaddress]{revtex4-2}
\usepackage{graphicx}
\usepackage{dcolumn}
\usepackage{bm}
\usepackage{hyperref}
\usepackage{orcidlink}

\usepackage{ulem}
\usepackage{xcolor}

\usepackage{soul}

\begin{document}

\preprint{APS/123-QED}

\title{Diagnosing Interstellar Magnetic Turbulence with TeV Pulsar Halos}

\author{Chao-Ming Li\orcidlink{0000-0002-9654-9123}}
\affiliation{School of Astronomy and Space Science, Nanjing University, Nanjing 210023, China}
\affiliation{Key Laboratory of Modern Astronomy and Astrophysics (Nanjing University), Ministry of Education, Nanjing 210023, China} 
\affiliation{Deutsches Elektronen Synchrotron (DESY), Platanenallee 6, D-15738 Zeuthen, Germany} 
\author{Ruo-Yu Liu\orcidlink{0000-0003-1576-0961}}%
\thanks{Corresponding author}
\email{ryliu@nju.edu.cn}
\affiliation{School of Astronomy and Space Science, Nanjing University, Nanjing 210023, China}
\affiliation{Key Laboratory of Modern Astronomy and Astrophysics (Nanjing University), Ministry of Education, Nanjing 210023, China} 
\author{Huirong Yan\orcidlink{0000-0003-2560-8066}}
\affiliation{Deutsches Elektronen Synchrotron (DESY), Platanenallee 6, D-15738 Zeuthen, Germany} 
\affiliation{Institut f{\"u}r Physik und Astronomie, Universit{\"a}t Potsdam, D-14476 Potsdam, Germany}


\date{\today}

\begin{abstract}
Interstellar magnetic field is essential in various astrophysical phenomena and processes. Pulsar halos are a recently discovered class of TeV gamma-ray sources formed by escaping electrons/positrons from pulsars. The morphology of the halo is regulated by the diffusion of those escaping particles, and hence carries information of the interstellar magnetic field. 
We suggest that the morphology of TeV pulsar halos can be used as a novel probe of the properties of interstellar magnetic field around the pulsar, such as the Alfv\'{e}nic Mach number and the mean direction. We establish a theoretical relation between these quantities and the observational features of the halo's morphology based on the anisotropic diffusion model, and show how X-ray observations of the pulsar halos can further improve the diagnosis of the magnetic field. 

\end{abstract}

\maketitle



{\it Introduction}---The magnetic field is a key ingredient of interstellar medium (ISM). It is deeply involved in various important astrophysical processes, such as cosmic-ray (CR) transport, the dynamics of molecular clouds and dust grains, and star formation. Many different tracers have been developed to measure the strength and direction of interstellar magnetic fields\cite{Beck_2015, Cho_2009, Chepurnov_2010, YL12, Pavaskar2023, Spangler_2001}. Currently, the primary method to diagnose the interstellar magnetic field is through the polarization measurements. Polarizations of molecular and atomic lines trace magnetic field directly in diffuse media, if available \cite{Crutcher2010, Zhang2020b}. The polarization of synchrotron radiation provides information about the perpendicular component\cite{Beck_2008} and inclination angle of magnetic fields \cite{Yuen2023, Malik2023}. In addition to polarization measurement, the Zeeman effect has been used to measure the field strength in molecular\cite{Fiebig89} or atomic\cite{Ching22} clouds. Faraday rotation can measure the parallel (line-of-sight) component of the large-scale, regular magnetic field \cite{Sokoloff_1998}.  


Pulsar halos are a new class of extended TeV gamma-ray sources recently discovered by the High Altitude Water Cherenkov Observatory (HAWC) and the Large High Altitude Air Shower Observatory (LHAASO) around three middle-aged ($\gtrsim 100\,$kyr) pulsars, the Geminga pulsar, the Monogem pulsar, and PSR~J0622+3749 \cite{hawc_2017,LHAASO_2021PhRvL}. A lot more candidates of pulsar halos have been revealed in the Galactic plane survey of the gamma-ray observatories\cite{HGPS_2018, LHAASO2024_catlog,HAWC_2020}. With a physical size of at least $20-30$\,pc, pulsar halos are distinct from pulsar wind nebulae (PWN)\cite{Giacinti2020}, which are usually quite compact ($\sim 0.1-1$\,pc) at the age of several hundred thousand years. Pulsar halos are believed to be produced by the inverse Compton (IC) scatterings of 
electrons/positrons (hereafter, we do not distinguish positrons from electrons for simplicity) escaping the PWN and diffusing in the ambient ISM\cite{Linden_2017,Liu_2022}. Although the IC radiation is not related to magnetic field, the spatial distribution of diffusing electrons is mainly determined by the properties of the background magnetic field, which plays a crucial role in shaping the halo morphology. The $\gamma$-ray emissions of halos are significantly detected at several tens of TeV, and for LHAASO~J0621+3755 the spectrum even goes up to 100\,TeV. These photons are emitted by electrons of energy $\gtrsim 100\,$TeV. At such high energies, the density of high-energy electrons within the halo is insufficient to amplify the turbulence via the streaming instability \cite{FangKun19, Malkov_2013,Mukhopadhyay22,Skilling_1975}. Therefore, the morphology of pulsar halos above a few tens of TeV mainly reflects properties of the interstellar magnetic field rather than self-triggered waves. 

There are no good tracers of magnetic field in the ambient ISM of PWNe except for radio and X-ray emissions, if available. For most pulsar halos and halo candidates, however, neither measurement has been reported. In this Letter, we attempt to develop a method of diagnosing the interstellar magnetic field using TeV gamma-ray observations of pulsar halos based on the anisotropic diffusion model\cite{Liu_2019, Yan_2022, Yan2025}. We will establish a relation of the inclination angle (denoted by $\phi$) of the mean magnetic field direction with respect to the observer's line-of-sight (LOS) and the Alfv{\'e}nic Mach number (denoted by $M_A$) of the ISM based on the morphology, or more specifically, the anisotropy of pulsar halos. We will also show that detection of pulsar halos in the X-ray band is crucial for breaking the degeneracy of these two quantities and help to reconstruct the three-dimensional structure as well as the strength of the mean magnetic field within the halo.

{\it Model}---The transport of particles within pulsar halos is dominated by diffusion, and the spatial distribution of particles $N(E_e,r)$ is subject to the property of the local magnetic turbulence. If the local turbulence is sub-Alfv{\'e}nic (i.e., $M_A<1$), the magnetic field has a local mean direction. Particles will  preferentially diffuse along the local mean direction, while the cross-field perpendicular diffusion would be suppressed by the gyro movements. The density distribution of particles is governed by the diffusion-loss equation 
\begin{equation}\label{eq:pde}
\begin{split}
\frac{\partial N}{\partial t}=&\frac{1}{r}\frac{\partial}{\partial r}\left(rD_\perp\frac{\partial N}{\partial r} \right)+D_\parallel\frac{\partial^2N}{\partial z^2}\\
&-\frac{\partial}{\partial E_e}\left(\dot{E}_eN\right)+Q(E_e)S(t)\delta(r)\delta(z),
\end{split}
\end{equation}
under the cylindrical coordinate system and assuming the mean direction aligns with $z$ direction. $\dot{E}_e$ is the cooling efficiency and $Q(E_e)S(t)\delta(r)\delta(z)$ is the injection term. Here we consider the symmetry of the system with respect to the $z-$axis, which is defined as the direction of the mean magnetic field. $D_\parallel$ and $D_\perp$ are the parallel diffusion coefficient and the perpendicular diffusion coefficient, which are related by the Alfv{\'e}nic Mach number($D_\perp = D_\parallel M_A^\xi$), where $\xi\simeq 4$ for Alfv{\'e}nic turbulence \citep{Ptuskin_1993, Yan2008}. It is mainly due to the scaling of weak turbulence from injection scale, the existence of which has been confirmed with in situ observations in space plasma \citep{Zhaosiqi2024}. $\xi$ becomes smaller if there is also non-negligible compressible component.  \citep{Maiti2022}. Modes identification in the local turbulence would give the valuable input \citep{Zhang2020}, which can be achieved through analysis of the radio emission from the same local volume with advanced technique such as the synchrotron polarization analysis (SPA+) technique \citep{Pavaskar2024}. As demonstrated in Ref.\citep{MakwanaYan2020}, the modes composition is largely subjected to the local driving mechanism of turbulence. 
Despite of the value of $\xi$, the parallel diffusion coefficient may be parameterized as $D_{\parallel}=D_0(E/{1\rm \,GeV})^{\delta}$ with $D_0$ being the value at 1\,GeV and $\delta$ being the energy dependence. 

By substituting $r'=r/M_{\rm A}^{\xi/2}$, the diffusion becomes isotropic under the new coordinate system $(r', z)$\cite{Fang2023}. The diffusion-loss equation may be further converted into spherical coordinate system with only radial dependence on $r'$, for which $N'(E_e, r')$ is given analytically \cite{Aharonian95, Liu20}\footnote{If the magnetic field within a pulsar halo is highly chaotic (i.e., $M_A \ge 1$), we have $D_\perp=D_\parallel$ and the diffusion of particles is roughly isotropic. The analytical solution can be used directly without coordinate transformation.}. The spatial distribution of particles in the original coordinate $(r,z)$ may be obtained by $N(E_e,r,z)=N'(E_e, \sqrt{z^2+r^2/M_{\rm A}^\xi})$. 
The solution implies that the electron distribution is akin to an ellipsoid, whose major axis aligns with the mean field direction. The 2D morphology of the $\gamma$-ray source produced by these electrons can be calculated by making the projection of the emission onto the plane of the sky \cite{Liu_2019,Yan_2022}.

\begin{figure}[htbp]
	\centering
	\includegraphics[width=0.5\textwidth]{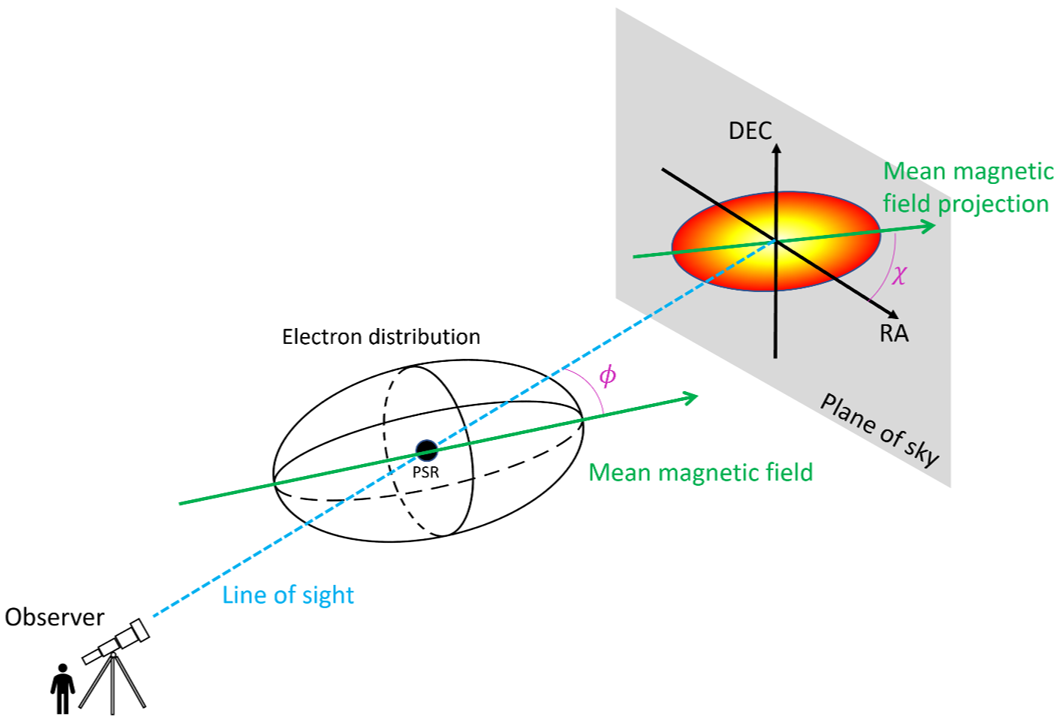}
	\caption{The 3D sketch of the pulsar halo and the corresponding TeV map on the plane of sky(grey shade). The dashed blue line is the direction of line-of-sight (LoS), while the solid black arrows are right ascension (RA) and declination (DEC). The coherent magnetic field lines are shown by the green arrows. The black ellipsoid denotes the distribution of electrons residing in such magnetic field. $\phi$ is  the inclination angle between LoS and mean magnetic field. $\chi$ is the angle between RA and the projection of magnetic field lines on the plane of sky.}
	\label{fig:sketch}
\end{figure}


\begin{figure}[htbp]
	\centering
	\includegraphics[width=0.5\textwidth]{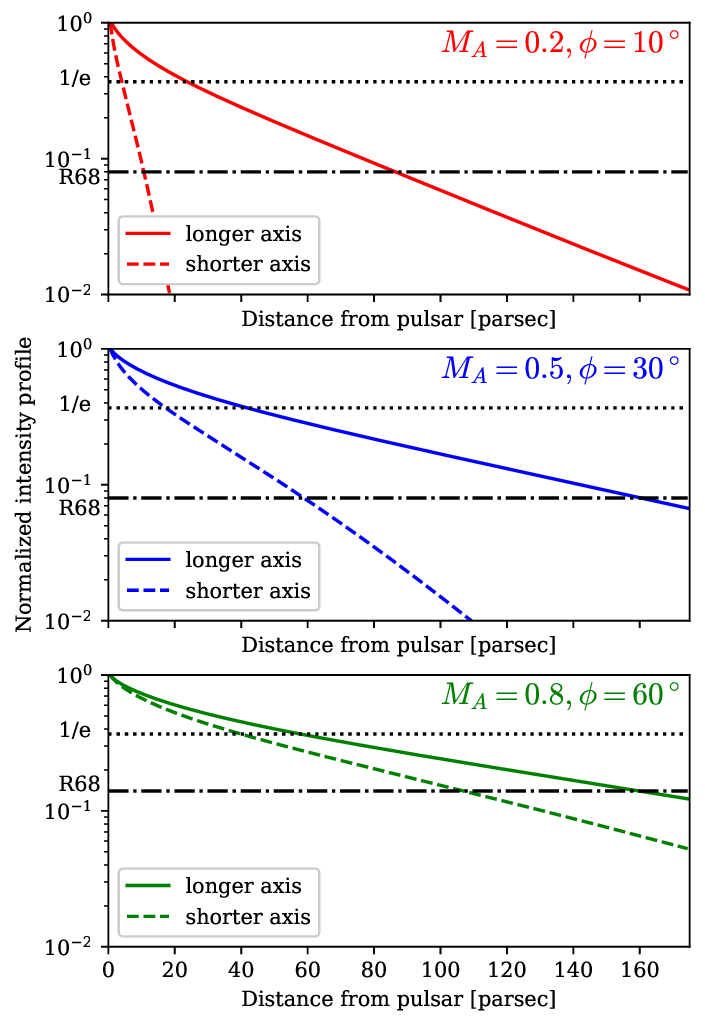}
	\caption{Normalized radial profile in 0.1 - 100\,TeV of the longer and shorter axes for three sets of $M_A$ and $\phi$ values, as texted in the plots. We use solid lines for longer axis and dashed lines for shorter axis. We also show the $1/e$ flux radius with dotted lines and $68\%$ containment radius ($R_{68}$) with dash-dotted lines. The spin-down luminosity of the test pulsar is assumed to be $10^{35}$\,erg/s. The injection spectrum is assumed to be a power-law distribution with a slope $-2$, followed by an exponential cutoff at energy 300 TeV. Note that the predicted profile is insensitive to these parameters. In the calculation, typical interstellar magnetic field strength $B=3\,\mu$G and parallel diffusion coefficient $D_0=10^{28} (E/\rm GeV)^{1/3} \, cm^2 s^{-1}$ are employed for calculation.}
	\label{fig:profile}
\end{figure}

The morphology of the source reflects the spatial distribution of electrons, which further reveals the properties of local interstellar magnetic field, such as the mean field direction and the Alfv{\'e}nic Mach number $M_{\rm A}$. It is straightforward to envisage that $\phi=0^\circ$ yields a circular projection of the source morphology, while $\phi>0^\circ$ 
produces an ellipselike morphology (see Fig.~\ref{fig:sketch})\footnote{If the pulsar is close to us, a slight asymmetry due to the distance effect will occur: the nearer side appears somewhat larger than the farther side, and the halo would appears a comet-like morphology\cite{Liu_2019}}. In the latter case, the longer axis follows the direction of the mean magnetic field projected onto the plane of the sky and the shorter axis is perpendicular to that direction. Therefore, we may obtain the azimuthal angle of the mean magnetic field $\chi$ (defined as the angle of the projected mean field with respect to the equatorial line) once the longer axis of the source is determined from observation. The ratio between the longer axis and the shorter axis also provides information about the magnetic field. Let us denote the length of the longer axis by $a$ and that of the shorter axis by $b$. Note that $a$ ($b$) is not exactly the major (minor) axis of an ellipse, as the source morphology is not a precise ellipse. Instead, $a$ and $b$ could be approximated by some typical length scales of the source. In Fig.~\ref{fig:profile}, we show the 1D intensity profile of the halo along the longer axis (solid curve) and the shorter axis (dashed curve) for different combinations of $M_{\rm A}$ and $\phi$. We may define the distance where the intensity drops to $1/e$ of the highest value (at the center of the source) as $a/2$ and $b/2$ in the longer axis and shorter axis, respectively. $a$ ($b$) could be also defined as the distance of the intersection of the $68\%$ containment contour on the longer (shorter) axis from the center. These typical lengths are related to the diffusion length of electrons parallel and perpendicular to the mean magnetic field direction, respectively, i.e., $\sqrt{D_\parallel t_{\rm c}}$ and $\sqrt{D_\perp t_{\rm c}}=M_{\rm A}^{\xi/2}\sqrt{D_\parallel t_{\rm c}}$, where $t_{\rm c}$ is the cooling timescale of electrons. Besides, due to the projection effect, $a$ should be also related to the inclination angle $\phi$.
Apparently, the axial ratio $a/b$ is expected to increase for a larger $\phi$ and reaches the maximum at $\phi=90^\circ$ for a fixed $M_{\rm A}$. On the other hand, a smaller $M_{\rm A}$ leads to a larger difference between $D_\parallel$ and $D_\perp$, and consequently a larger axial ratio $a/b$ (except $\phi=0$). The diffusion coefficient is expected to be canceled in the axial ratio.  We have also tested different spectral shape of injection electrons, and found that they have no influence on the axial ratio.

\begin{figure}[htbp]
	\centering	\includegraphics[width=0.5\textwidth]{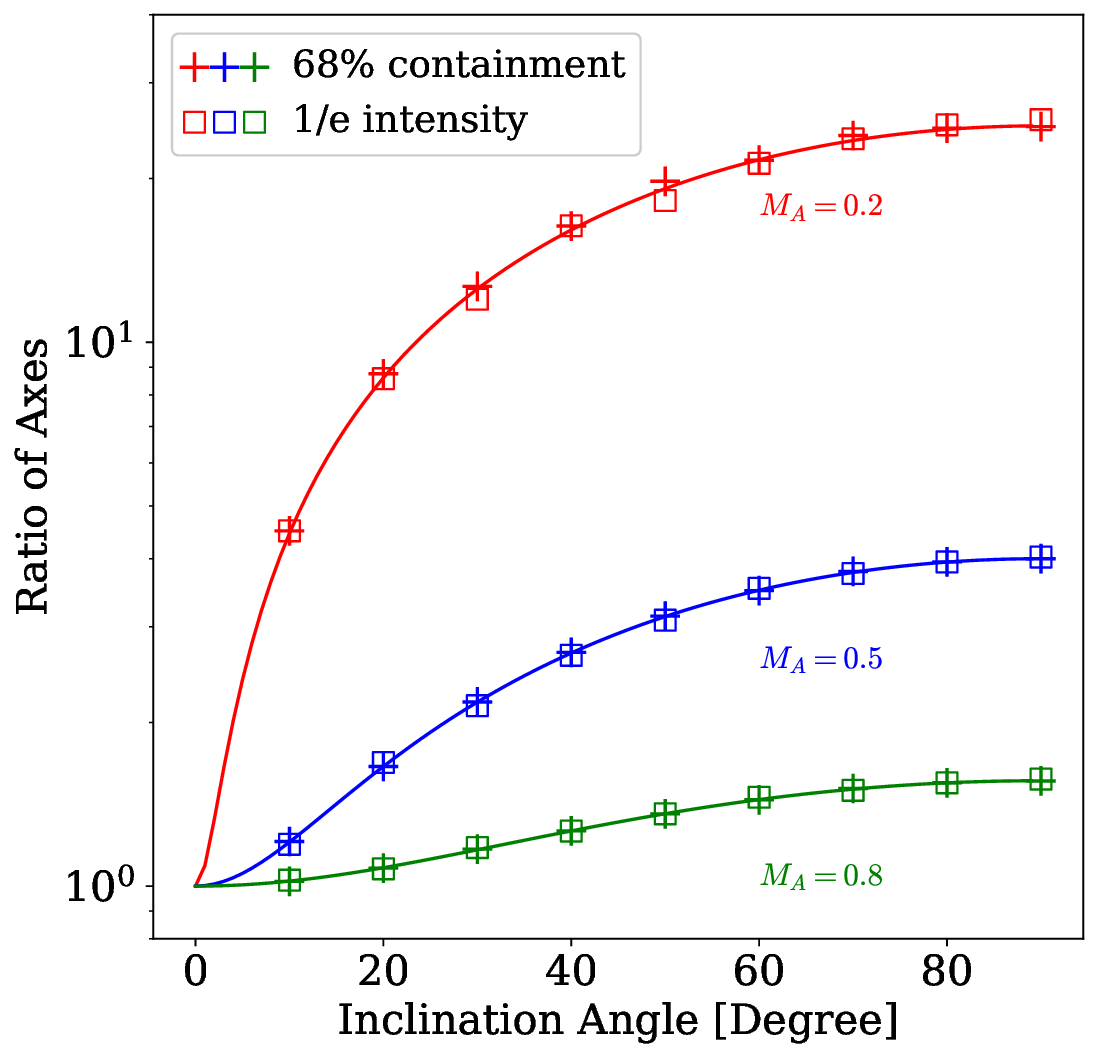}
	\caption{The ratio of longer axis and shorter axis $(a/b)$ for different Alfv\'{e}nic Mach number and inclination angles. Crosses and squares represent the outcomes of numerical calculations for two different definitions of the axes as marked in the legend, whereas the solid curves correspond to the analytical results formulated in Eq. (\ref{eq:relation}). The parameters of the pulsar are the same as Fig.~\ref{fig:profile}.}
	\label{fig:axes_ratio}
\end{figure}

{\it Result.}---No matter how we define $a$ and $b$, the ratio $a/b$ is the same for a given combination of $M_{\rm A}$ and $\phi$. We may derive a quantitative dependence of $a/b$ on $M_{\rm A}$ and $\phi$. As detailed in Supplemental Material\footnote{\url{https://journals.aps.org/prd/abstract/10.1103/vt3s-rbj1\#supplemental}}, a simple relation can be given by
\begin{equation}\label{eq:relation}
    \frac{a}{b} = \sqrt{\frac{\sin^2\phi}{M_A^\xi} + \cos^2\phi},
\end{equation}
as long as the source distance $d$ is much larger than $a$ and $b$. In the following calculation, we assume that the Alfv{\'e}nic mode dominates the turbulence and hence take $\xi=4$. In Fig.~\ref{fig:axes_ratio}, we show the axial ratio as a function of the inclination angle $\phi$ for $M_{\rm A}=0.2$ (red), 0.5 (blue), and 0.8 (green). The solid curves in the figure show the relation given by Eq.~(\ref{eq:relation}). Squares and crosses show the axial ratios obtained numerically based on the definitions of the $1/e$ maximum intensity and the 68\% flux containment radius, respectively, under different combinations of $\phi$ and $M_{\rm A}$. We see the simple analytical relation given by Eq.~(\ref{eq:relation}) and our numerical results match each other very well. Also, the axial ratios numerically obtained with the two definitions are consistent with each other. 

\begin{figure}[htbp]
	\centering
	\includegraphics[width=0.5\textwidth]{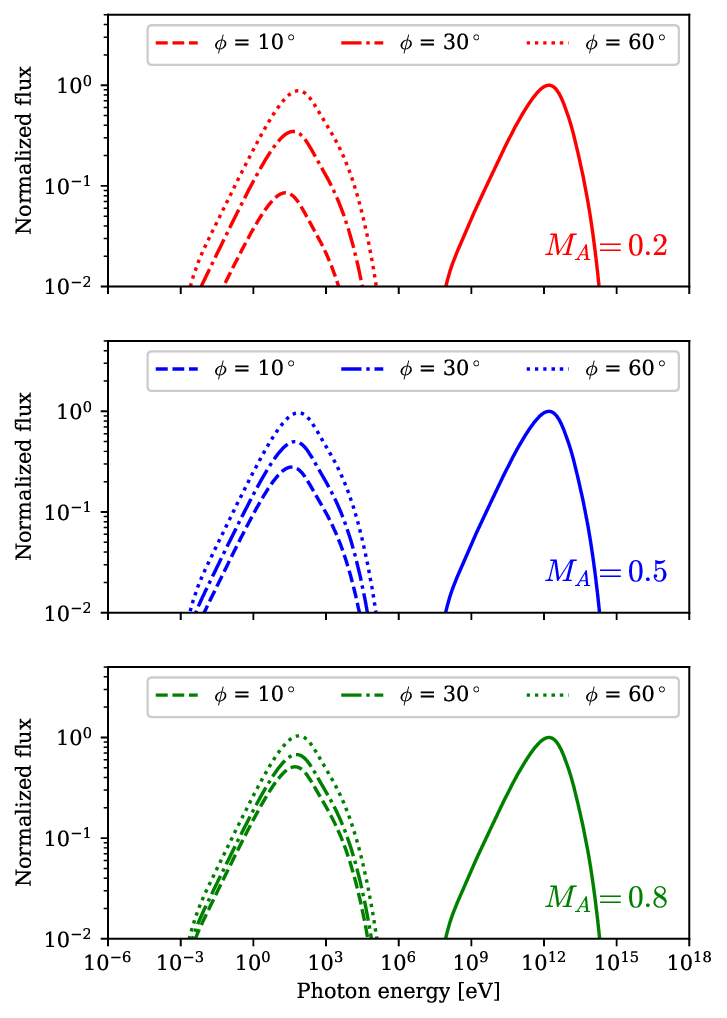}
	\caption{Predicted SED with different Alfv\'{e}nic Mach number and inclination angles. Red, blue and green colors represent $M_A$ = 0.2, 0.5 and 0.8 respectively. The IC peak flux is normalized at unity. Dashed, dash-dotted and dotted linestyles represent inclination angle $\phi=10^\circ, 30^\circ$ and $60^\circ$ respectively. The solid lines are inverse Compton emission, which remain the same despite varying inclination angles and Alfv\'{e}nic Mach numbers. The parameters of the pulsar are the same as Fig.~\ref{fig:profile}. }
	\label{fig:syn}
\end{figure}

Therefore, based on measurements of (candidate) pulsar halos by instruments such as LHAASO, HAWC, HESS, etc, we may derive the azimuthal angle $\chi$ of the mean magnetic field, and obtain a relation between the inclination angle $\phi$ and the Alfv{\'e}nic Mach number $M_{\rm A}$. If we want to further break the degeneracy between $\phi$ and $M_{\rm A}$, observations at other wavelengths are needed, particularly radio and X-ray observations. Radio polarization observations can provide information of the turbulent magnetic field with various methods such as the aforementioned SPA+ technique \cite{Pavaskar2024}. We here mainly discuss the role of X-ray observations. Tens of TeV photons are mainly produced by $\sim 100$\,TeV electrons via the IC scattering off cosmic microwave background photons with a minor contribution from the interstellar infrared radiation background, and
the same electrons will also produce X-ray emission via the synchrotron radiation in the typical interstellar magnetic field of $B\sim$ a few $\mu$G, i.e., $\epsilon_{\rm syn}=2\,(E_e/100{\rm TeV})^2(B/3\mu{\rm G})\sin\alpha\,$keV, where $\alpha$ is the pitch angle between the magnetic field direction and the moving direction of emitting electrons. The IC radiation power $P_{\rm IC}$ depends on the energy density of the target radiation field, which is basically known. On the other hand, the synchrotron radiation power is related to the magnetic field energy density and the pitch angle $\alpha$. The latter is equal to the inclination angle $\phi$ if we only consider the mean magnetic field. However, more strictly speaking, the average pitch angle is not exactly equal to the inclination angle, because the field line is perturbed by the turbulence. Ref.\cite{Lazarian_2012} found that the average cosine of the angle between the local field and the mean field can be given by $\cos^2\delta=(1+M_{\rm A}^2)^{-1}$. The average cosine of the pitch angle can then be estimated by $\cos^2\alpha=(1+M_{\rm A}^2)^{-1}\cos^2\phi$. The synchrotron radiation power is then proportional to $P_{\rm syn}\propto \sin^2\alpha(B^2/8\pi)$. Therefore, if the X-ray emission of pulsar halos can be detected, we may obtain another independent constraint on the combination of $M_{\rm A}$ and $\phi$ based on the ratio between the synchrotron flux and the IC flux, or via the spectral energy distribution (SED) modeling, if the interstellar magnetic field strength $B$ is also known. There are some ways of evaluating the value of $B$. As a rough estimation, it could be 
assumed as a typical value such as $5\,\mu$G or the value given by models (e.g., Ref.\cite{Jansson2012, Unger2024}). Alternatively, $B$ may be obtained according to the measured size of the shorter axis of the source $b$. The corresponding physical size of $b$ may indicate the diffusion length of electrons perpendicular to the mean field direction before cooling, i.e., $b \approx 2\sqrt{M_A^4D_\parallel t_{\rm c}}$, where $t_{\rm c}$ is the cooling timescale and related to $B$, so that the value of $B$ may be solved together with $M_A$ and $\phi$. In Fig.~\ref{fig:syn}, we show the expected broadband SED of pulsar halos with different $M_{\rm A}$ and $\phi$. We see that the X-ray flux changes significantly with different combination of $M_{\rm A}$ and $\phi$, while the TeV flux keeps the same.  

Note that, the X-ray polarization measurements could also be helpful to disentangle the degeneracy between $\phi$ and $M_A$ in principle, but it would be challenging for the current generation of X-ray polarization instruments such as IXPE, because the size of the halo is much bigger than their field of view, and the X-ray flux of pulsar halos is generally weak\citep{Liu2019b, Khokhriakova2024, Manconi2024} and below the detection limit. 

{\it Discussion.} --- Our method is applicable to pulsar halos of homogeneous diffusion properties with magnetic coherence length $L \gtrsim$ the halo's size, which is typically a few tens to a few hundreds of parsecs\cite{Falceta_2015, Beck_2016}. It corresponds to a sub-Alfv{\'e}nic turbulence ($M_A < 1$), where the mean direction of the magnetic field can be defined over the spatial scale of the halo (or at least a large fraction of it). If the magnetic field within the halo is very chaotic (i.e. $M_A>1$) or the coherence length $L$ is significantly smaller than the halo size, particles would diffuse isotropically, resulting in a more or less spherical morphology. In this case, Eq.~(\ref{eq:relation}) will no longer be applicable. 

Recently, HAWC Collaboration\citep{HAWC_2024} published their updated measurements on halos of the Geminga pulsar and the Monogem pulsar, and found asymmetric morphology of the halos at a significance level of $\gtrsim 2\sigma$, which corroborates our anisotropic diffusion framework. 
In the future, with accumulated data on pulsar halos, the anisotropic nature could be more evident. And if TeV pulsar halos or the candidates can be measured by X-ray and/or radio instruments with large fields of view, it will facilitate our understanding of the interstellar magnetic field alongside TeV observations.

\appendix


\bibliographystyle{apsrev}
\bibliography{ms}

@ARTICLE{Liu2019b,
       author = {{Liu}, Ruo-Yu and {Ge}, Chong and {Sun}, Xiao-Na and {Wang}, Xiang-Yu},
        title = "{Constraining the Magnetic Field in the TeV Halo of Geminga with X-Ray Observations}",
      journal = {apj},
         year = 2019,
        month = apr,
       volume = {875},
       number = {2},
          eid = {149},
        pages = {149},
          doi = {10.3847/1538-4357/ab125c},
archivePrefix = {arXiv},
       eprint = {1904.11438},
 primaryClass = {astro-ph.HE},
       adsurl = {https://ui.adsabs.harvard.edu/abs/2019ApJ...875..149L}
}

@ARTICLE{Manconi2024,
       author = {{Manconi}, Silvia and {Woo}, Jooyun and {Shang}, Ruo-Yu and {Krivonos}, Roman and {Tang}, Claudia and {Di Mauro}, Mattia and {Donato}, Fiorenza and {Mori}, Kaya and {Hailey}, Charles J.},
        title = "{Geminga's pulsar halo: An X-ray view}",
      journal = {aap},
         year = 2024,
        month = sep,
       volume = {689},
          eid = {A326},
        pages = {A326},
          doi = {10.1051/0004-6361/202450242},
archivePrefix = {arXiv},
       eprint = {2403.10902},
 primaryClass = {astro-ph.HE},
       adsurl = {https://ui.adsabs.harvard.edu/abs/2024A&A...689A.326M}
}

@ARTICLE{Khokhriakova2024,
       author = {{Khokhriakova}, A. and {Becker}, W. and {Ponti}, G. and {Sasaki}, M. and {Li}, B. and {Liu}, R.-Y.},
        title = "{Searching for X-ray counterparts of degree-wide TeV halos around middle-aged pulsars with SRG/eROSITA}",
      journal = {aap},
         year = 2024,
        month = mar,
       volume = {683},
          eid = {A180},
        pages = {A180},
          doi = {10.1051/0004-6361/202347311},
archivePrefix = {arXiv},
       eprint = {2310.10454},
 primaryClass = {astro-ph.HE},
       adsurl = {https://ui.adsabs.harvard.edu/abs/2024A&A...683A.180K}
}

@ARTICLE{Yan2025,
       author = {{Yan}, Kai and {Wu}, Sha and {Liu}, Ruo-Yu},
        title = "{Anisotropic Diffusion of e$^{{\ensuremath{\pm}}}$ in Pulsar Halos over Multiple Coherence of Magnetic Fields}",
      journal = {\apj},
         year = 2025,
        month = jul,
       volume = {987},
       number = {1},
          eid = {19},
        pages = {19},
          doi = {10.3847/1538-4357/add6a4},
archivePrefix = {arXiv},
 primaryClass = {astro-ph.HE},
       adsurl = {https://ui.adsabs.harvard.edu/abs/2025ApJ...987...19Y}
}

@ARTICLE{Fiebig89,
       author = {{Fiebig}, D. and {Guesten}, R.},
        title = "{Strong magnetic fields in interstellar H2O maser clumps.}",
      journal = {Astron. Astrophys.},
         year = 1989,
        month = apr,
       volume = {214},
        pages = {333-338},
       adsurl = {https://ui.adsabs.harvard.edu/abs/1989A&A...214..333F}
}

@ARTICLE{Ching22,
       author = {{Ching}, T. -C. and {Li}, D. and {Heiles}, C. and {Li}, Z. -Y. and {Qian}, L. and {Yue}, Y.~L. and {Tang}, J. and {Jiao}, S.~H.},
        title = "{An early transition to magnetic supercriticality in star formation}",
      journal = {Nature(London)},
         year = 2022,
        month = jan,
       volume = {601},
       number = {7891},
        pages = {49-52},
          doi = {10.1038/s41586-021-04159-x},
archivePrefix = {arXiv},
 primaryClass = {astro-ph.GA},
       adsurl = {https://ui.adsabs.harvard.edu/abs/2022Natur.601...49C}
}

@ARTICLE{Pavaskar2023,
       author = {{Pavaskar}, Parth and {Yan}, Huirong and {Cho}, Jungyeon},
        title = "{Magnetic field measurement from the Davis-Chandrasekhar-Fermi method employed with atomic alignment}",
      journal = {Mon. Not. R. Astron. Soc.},
         year = 2023,
        month = jul,
       volume = {523},
       number = {1},
        pages = {1056-1066},
          doi = {10.1093/mnras/stad1237},
archivePrefix = {arXiv},
 primaryClass = {astro-ph.IM},
       adsurl = {https://ui.adsabs.harvard.edu/abs/2023MNRAS.523.1056P}
}

@ARTICLE{Maiti2022,
       author = {{Maiti}, Snehanshu and {Makwana}, Kirit and {Zhang}, Heshou and {Yan}, Huirong},
        title = "{Cosmic-ray Transport in Magnetohydrodynamic Turbulence}",
      journal = {Astrophys. J.},
         year = 2022,
        month = feb,
       volume = {926},
       number = {1},
          eid = {94},
        pages = {94},
          doi = {10.3847/1538-4357/ac46c8},
archivePrefix = {arXiv},
 primaryClass = {astro-ph.HE},
       adsurl = {https://ui.adsabs.harvard.edu/abs/2022ApJ...926...94M}
}

@ARTICLE{YL12,
       author = {{Yan}, Huirong and {Lazarian}, A.},
        title = "{Tracing magnetic fields with ground state alignment}",
      journal = {J. Quant. Spectrosc. Radiat. Transfer.},
         year = 2012,
        month = aug,
       volume = {113},
       number = {12},
        pages = {1409-1428},
          doi = {10.1016/j.jqsrt.2012.03.027},
archivePrefix = {arXiv},
 primaryClass = {astro-ph.GA},
       adsurl = {https://ui.adsabs.harvard.edu/abs/2012JQSRT.113.1409Y}
}

@ARTICLE{Zhaosiqi2024,
       author = {{Zhao}, Siqi and {Yan}, Huirong and {Liu}, Terry Z. and {Yuen}, Ka Ho and {Wang}, Huizi},
        title = "{Identification of the weak-to-strong transition in Alfv{\'e}nic turbulence from space plasma}",
      journal = {Nat. Astron.},
         year = 2024,
        month = jun,
       volume = {8},
        pages = {725-731},
          doi = {10.1038/s41550-024-02249-0},
archivePrefix = {arXiv},
 primaryClass = {astro-ph.SR},
       adsurl = {https://ui.adsabs.harvard.edu/abs/2024NatAs...8..725Z}
}

@ARTICLE{MakwanaYan2020,
       author = {{Makwana}, K.~D. and {Yan}, Huirong},
        title = "{Properties of Magnetohydrodynamic Modes in Compressively Driven Plasma Turbulence}",
      journal = {Phys. Rev. X.},
         year = 2020,
        month = jul,
       volume = {10},
       number = {3},
          eid = {031021},
        pages = {031021},
          doi = {10.1103/PhysRevX.10.031021},
archivePrefix = {arXiv},
 primaryClass = {physics.plasm-ph},
       adsurl = {https://ui.adsabs.harvard.edu/abs/2020PhRvX..10c1021M}
}

@ARTICLE{Pavaskar2024,
       author = {{Pavaskar}, Parth and {Yuen}, Ka Ho and {Yan}, Huirong and {Malik}, Sunil},
        title = "{Diagnostics of Magnetohydrodynamic Modes in the Interstellar Medium through Synchrotron Polarization Statistics}",
      journal = {Astrophys. J.},
         year = 2024,
        month = aug,
       volume = {971},
       number = {1},
          eid = {58},
        pages = {58},
          doi = {10.3847/1538-4357/ad5af5},
archivePrefix = {arXiv},
 primaryClass = {astro-ph.GA},
       adsurl = {https://ui.adsabs.harvard.edu/abs/2024ApJ...971...58P}
}

@ARTICLE{Zhang2020,
       author = {{Zhang}, Heshou and {Chepurnov}, Alexey and {Yan}, Huirong and {Makwana}, Kirit and {Santos-Lima}, Reinaldo and {Appleby}, Sarah},
        title = "{Identification of plasma modes in Galactic turbulence with synchrotron polarization}",
      journal = {Nat. Astron.},
         year = 2020,
        month = may,
       volume = {4},
        pages = {1001-1008},
          doi = {10.1038/s41550-020-1093-4},
archivePrefix = {arXiv},
 primaryClass = {physics.plasm-ph},
       adsurl = {https://ui.adsabs.harvard.edu/abs/2020NatAs...4.1001Z}
}

@ARTICLE{Zhang2020b,
       author = {{Zhang}, Heshou and {Gangi}, Manuele and {Leone}, Francesco and {Taylor}, Andrew and {Yan}, Huirong},
        title = "{Discovery of Ground-state Absorption Line Polarization and Sub-Gauss Magnetic Field in the Post-AGB Binary System 89 Her}",
      journal = {Astrophys. J. Lett.},
         year = 2020,
        month = oct,
       volume = {902},
       number = {1},
          eid = {L7},
        pages = {L7},
          doi = {10.3847/2041-8213/abb8e1},
archivePrefix = {arXiv},
 primaryClass = {astro-ph.GA},
       adsurl = {https://ui.adsabs.harvard.edu/abs/2020ApJ...902L...7Z}
}

@ARTICLE{Malik2023,
       author = {{Malik}, Sunil and {Yuen}, Ka Ho and {Yan}, Huirong},
        title = "{Diagnosis of 3D magnetic field and mode composition in MHD turbulence with Y-parameter}",
      journal = {Mon. Not. R. Astron. Soc.},
         year = 2023,
        month = oct,
       volume = {524},
       number = {4},
        pages = {6102-6113},
          doi = {10.1093/mnras/stad2225},
archivePrefix = {arXiv},
 primaryClass = {astro-ph.GA},
       adsurl = {https://ui.adsabs.harvard.edu/abs/2023MNRAS.524.6102M}
}

@ARTICLE{Crutcher2010,
       author = {{Crutcher}, Richard M. and {Wandelt}, Benjamin and {Heiles}, Carl and {Falgarone}, Edith and {Troland}, Thomas H.},
        title = "{Magnetic Fields in Interstellar Clouds from Zeeman Observations: Inference of Total Field Strengths by Bayesian Analysis}",
      journal = {Astrophys. J.},
         year = 2010,
        month = dec,
       volume = {725},
       number = {1},
        pages = {466-479},
          doi = {10.1088/0004-637X/725/1/466},
       adsurl = {https://ui.adsabs.harvard.edu/abs/2010ApJ...725..466C}
}

@ARTICLE{Yuen2023,
       author = {{Yuen}, Ka Ho and {Yan}, Huirong and {Lazarian}, Alex},
        title = "{Anomalous compressible mode generation by global frame projections of pure Alfven mode}",
      journal = {Mon. Not. R. Astron. Soc.},
         year = 2023,
        month = may,
       volume = {521},
       number = {1},
        pages = {530-545},
          doi = {10.1093/mnras/stad287},
archivePrefix = {arXiv},
 primaryClass = {astro-ph.GA},
       adsurl = {https://ui.adsabs.harvard.edu/abs/2023MNRAS.521..530Y}
}

@ARTICLE{Aharonian95,
       author = {{Aharonian}, F.~A. and {Atoyan}, A.~M. and {Voelk}, H.~J.},
        title = "{High energy electrons and positrons in cosmic rays as an indicator of the existence of a nearby cosmic tevatron}",
      journal = {Astron. Astrophys.},
         year = 1995,
        month = feb,
       volume = {294},
        pages = {L41-L44},
       adsurl = {https://ui.adsabs.harvard.edu/abs/1995A&A...294L..41A}
}

@ARTICLE{Liu20,
       author = {{Liu}, Ruo-Yu and {Yan}, Huirong},
        title = "{On the unusually large spatial extent of the TeV nebula HESS J1825-137: implication from the energy-dependent morphology}",
      journal = {Mon. Not. R. Astron. Soc.},
         year = 2020,
        month = may,
       volume = {494},
       number = {2},
        pages = {2618-2627},
          doi = {10.1093/mnras/staa911},
archivePrefix = {arXiv},
 primaryClass = {astro-ph.HE},
       adsurl = {https://ui.adsabs.harvard.edu/abs/2020MNRAS.494.2618L}
}

@ARTICLE{Fangkun19,
       author = {{Fang}, Kun and {Bi}, Xiao-Jun and {Yin}, Peng-Fei},
        title = "{Possible origin of the slow-diffusion region around Geminga}",
      journal = {Mon. Not. R. Astron. Soc.},
         year = 2019,
        month = sep,
       volume = {488},
       number = {3},
        pages = {4074-4080},
          doi = {10.1093/mnras/stz1974},
archivePrefix = {arXiv},
 primaryClass = {astro-ph.HE},
       adsurl = {https://ui.adsabs.harvard.edu/abs/2019MNRAS.488.4074F}
}

@ARTICLE{Mukhopadhyay22,
       author = {{Mukhopadhyay}, Payel and {Linden}, Tim},
        title = "{Self-generated cosmic-ray turbulence can explain the morphology of TeV halos}",
      journal = {Phys. Rev. D.},
         year = 2022,
        month = jun,
       volume = {105},
       number = {12},
          eid = {123008},
        pages = {123008},
          doi = {10.1103/PhysRevD.105.123008},
archivePrefix = {arXiv},
 primaryClass = {astro-ph.HE},
       adsurl = {https://ui.adsabs.harvard.edu/abs/2022PhRvD.105l3008M}
}

@ARTICLE{Jansson2012,
       author = {{Jansson}, Ronnie and {Farrar}, Glennys R.},
        title = "{A New Model of the Galactic Magnetic Field}",
      journal = {Astrophys. J.},
         year = 2012,
        month = sep,
       volume = {757},
       number = {1},
          eid = {14},
        pages = {14},
          doi = {10.1088/0004-637X/757/1/14},
archivePrefix = {arXiv},
 primaryClass = {astro-ph.GA},
       adsurl = {https://ui.adsabs.harvard.edu/abs/2012ApJ...757...14J}
}

@ARTICLE{Unger2024,
       author = {{Unger}, Michael and {Farrar}, Glennys R.},
        title = "{The Coherent Magnetic Field of the Milky Way}",
      journal = {Astrophys. J.},
         year = 2024,
        month = jul,
       volume = {970},
       number = {1},
          eid = {95},
        pages = {95},
          doi = {10.3847/1538-4357/ad4a54},
archivePrefix = {arXiv},
 primaryClass = {astro-ph.GA},
       adsurl = {https://ui.adsabs.harvard.edu/abs/2024ApJ...970...95U}
}

@article{Fang2023,
  title = {Effect of the magnetic field correlation length on the gamma-ray pulsar halo morphology under anisotropic diffusion},
  author = {Fang, Kun and Hu, Hong-Bo and Bi, Xiao-Jun and Chen, En-Sheng},
  journal = {Phys. Rev. D.},
  volume = {108},
  issue = {2},
  pages = {023017},
  numpages = {9},
  year = {2023},
  month = {Jul},
  publisher = {American Physical Society},
  doi = {10.1103/PhysRevD.108.023017},
}

@ARTICLE{Yan2008,
	author = {{Yan}, Huirong and {Lazarian}, A.},
	title = "{Cosmic-Ray Propagation: Nonlinear Diffusion Parallel and Perpendicular to Mean Magnetic Field}",
	journal = {Astrophys. J.},
	year = 2008,
	month = feb,
	volume = {673},
	number = {2},
	pages = {942-953},
	doi = {10.1086/524771},
	archivePrefix = {arXiv},
	primaryClass = {astro-ph},
	adsurl = {https://ui.adsabs.harvard.edu/abs/2008ApJ...673..942Y}
}

@ARTICLE{Giacinti2020,
	author = {{Giacinti}, G. and {Mitchell}, A.~M.~W. and {L{'o}pez-Coto}, R. and {Joshi}, V. and {Parsons}, R.~D. and {Hinton}, J.~A.},
	title = "{Halo fraction in TeV-bright pulsar wind nebulae}",
	journal = {Astron. Astrophys.},
	year = 2020,
	month = apr,
	volume = {636},
	eid = {A113},
	pages = {A113},
	doi = {10.1051/0004-6361/201936505},
	archivePrefix = {arXiv},
	primaryClass = {astro-ph.HE},
	adsurl = {https://ui.adsabs.harvard.edu/abs/2020A&A...636A.113G}
}

@article{HAWC_2020,
	title = {Multiple Galactic Sources with Emission Above 56 TeV Detected by HAWC},
	author = {Abeysekara, A. U. and Albert, A. and Alfaro, R. and Angeles Camacho, J. R. and Arteaga-Vel\'azquez, J. C. and Arunbabu, K. P. and Avila Rojas, D. and Ayala Solares, H. A. and Baghmanyan, V. and Belmont-Moreno, E. and BenZvi, S. Y. and Brisbois, C. and Caballero-Mora, K. S. and Capistr\'an, T. and Carrami\~nana, A. and Casanova, S. and Cotti, U. and Cotzomi, J. and Couti\~no de Le\'on, S. and De la Fuente, E. and de Le\'on, C. and Dichiara, S. and Dingus, B. L. and DuVernois, M. A. and D\'{i}az-V\'elez, J. C. and Ellsworth, R. W. and Engel, K. and Espinoza, C. and Fleischhack, H. and Fraija, N. and Galv\'an-G\'amez, A. and Garcia, D. and Garc\'{i}a-Gonz\'alez, J. A. and Garfias, F. and Gonz\'alez, M. M. and Goodman, J. A. and Harding, J. P. and Hernandez, S. and Hinton, J. and Hona, B. and Huang, D. and Hueyotl-Zahuantitla, F. and H\"untemeyer, P. and Iriarte, A. and Jardin-Blicq, A. and Joshi, V. and Kaufmann, S. and Kieda, D. and Lara, A. and Lee, W. H. and Le\'on Vargas, H. and Linnemann, J. T. and Longinotti, A. L. and Luis-Raya, G. and Lundeen, J. and L\'opez-Coto, R. and Malone, K. and Marinelli, S. S. and Martinez, O. and Martinez-Castellanos, I. and Mart\'{i}nez-Castro, J. and Mart\'{i}nez-Huerta, H. and Matthews, J. A. and Miranda-Romagnoli, P. and Morales-Soto, J. A. and Moreno, E. and Mostaf\'a, M. and Nayerhoda, A. and Nellen, L. and Newbold, M. and Nisa, M. U. and Noriega-Papaqui, R. and Peisker, A. and P\'erez-P\'erez, E. G. and Pretz, J. and Ren, Z. and Rho, C. D. and Rivi\`ere, C. and Rosa-Gonz\'alez, D. and Rosenberg, M. and Ruiz-Velasco, E. and Salesa Greus, F. and Sandoval, A. and Schneider, M. and Schoorlemmer, H. and Sinnis, G. and Smith, A. J. and Springer, R. W. and Surajbali, P. and Tabachnick, E. and Tanner, M. and Tibolla, O. and Tollefson, K. and Torres, I. and Torres-Escobedo, R. and Villase\~nor, L. and Weisgarber, T. and Wood, J. and Yapici, T. and Zhang, H. and Zhou, H.},
	collaboration = {HAWC Collaboration},
	journal = {Phys. Rev. Lett.},
	volume = {124},
	issue = {2},
	pages = {021102},
	numpages = {7},
	year = {2020},
	month = {Jan},
	publisher = {American Physical Society},
	doi = {10.1103/PhysRevLett.124.021102},
}

@ARTICLE{HGPS_2018,
	author = {{H.~E.~S.~S. Collaboration} and {Abdalla}, H. and {Abramowski}, A. and {Aharonian}, F. and {Ait Benkhali}, F. and {Ang{"u}ner}, E.~O. and {Arakawa}, M. and {Arrieta}, M. and {Aubert}, P. and {Backes}, M. and {Balzer}, A. and {Barnard}, M. and {Becherini}, Y. and {Becker Tjus}, J. and {Berge}, D. and {Bernhard}, S. and {Bernl{"o}hr}, K. and {Blackwell}, R. and {B{"o}ttcher}, M. and {Boisson}, C. and {Bolmont}, J. and {Bonnefoy}, S. and {Bordas}, P. and {Bregeon}, J. and {Brun}, F. and {Brun}, P. and {Bryan}, M. and {B{"u}chele}, M. and {Bulik}, T. and {Capasso}, M. and {Carrigan}, S. and {Caroff}, S. and {Carosi}, A. and {Casanova}, S. and {Cerruti}, M. and {Chakraborty}, N. and {Chaves}, R.~C.~G. and {Chen}, A. and {Chevalier}, J. and {Colafrancesco}, S. and {Condon}, B. and {Conrad}, J. and {Davids}, I.~D. and {Decock}, J. and {Deil}, C. and {Devin}, J. and {deWilt}, P. and {Dirson}, L. and {Djannati-Ata{"\i}}, A. and {Domainko}, W. and {Donath}, A. and {Drury}, L.~O. 'C. and {Dutson}, K. and {Dyks}, J. and {Edwards}, T. and {Egberts}, K. and {Eger}, P. and {Emery}, G. and {Ernenwein}, J. -P. and {Eschbach}, S. and {Farnier}, C. and {Fegan}, S. and {Fernandes}, M.~V. and {Fiasson}, A. and {Fontaine}, G. and {F{"o}rster}, A. and {Funk}, S. and {F{"u}{ss}ling}, M. and {Gabici}, S. and {Gallant}, Y.~A. and {Garrigoux}, T. and {Gast}, H. and {Gat{'e}}, F. and {Giavitto}, G. and {Giebels}, B. and {Glawion}, D. and {Glicenstein}, J.~F. and {Gottschall}, D. and {Grondin}, M. -H. and {Hahn}, J. and {Haupt}, M. and {Hawkes}, J. and {Heinzelmann}, G. and {Henri}, G. and {Hermann}, G. and {Hinton}, J.~A. and {Hofmann}, W. and {Hoischen}, C. and {Holch}, T.~L. and {Holler}, M. and {Horns}, D. and {Ivascenko}, A. and {Iwasaki}, H. and {Jacholkowska}, A. and {Jamrozy}, M. and {Jankowsky}, D. and {Jankowsky}, F. and {Jingo}, M. and {Jouvin}, L. and {Jung-Richardt}, I. and {Kastendieck}, M.~A. and {Katarzy{'n}ski}, K. and {Katsuragawa}, M. and {Katz}, U. and {Kerszberg}, D. and {Khangulyan}, D. and {Kh{'e}lifi}, B. and {King}, J. and {Klepser}, S. and {Klochkov}, D. and {Klu{'z}niak}, W. and {Komin}, Nu. and {Kosack}, K. and {Krakau}, S. and {Kraus}, M. and {Kr{"u}ger}, P.~P. and {Laffon}, H. and {Lamanna}, G. and {Lau}, J. and {Lees}, J. -P. and {Lefaucheur}, J. and {Lemi{`e}re}, A. and {Lemoine-Goumard}, M. and {Lenain}, J. -P. and {Leser}, E. and {Lohse}, T. and {Lorentz}, M. and {Liu}, R. and {L{'o}pez-Coto}, R. and {Lypova}, I. and {Marandon}, V. and {Malyshev}, D. and {Marcowith}, A. and {Mariaud}, C. and {Marx}, R. and {Maurin}, G. and {Maxted}, N. and {Mayer}, M. and {Meintjes}, P.~J. and {Meyer}, M. and {Mitchell}, A.~M.~W. and {Moderski}, R. and {Mohamed}, M. and {Mohrmann}, L. and {Mor{r{a}}}, K. and {Moulin}, E. and {Murach}, T. and {Nakashima}, S. and {de Naurois}, M. and {Ndiyavala}, H. and {Niederwanger}, F. and {Niemiec}, J. and {Oakes}, L. and {O'Brien}, P. and {Odaka}, H. and {Ohm}, S. and {Ostrowski}, M. and {Oya}, I. and {Padovani}, M. and {Panter}, M. and {Parsons}, R.~D. and {Paz Arribas}, M. and {Pekeur}, N.~W. and {Pelletier}, G. and {Perennes}, C. and {Petrucci}, P. -O. and {Peyaud}, B. and {Piel}, Q. and {Pita}, S. and {Poireau}, V. and {Poon}, H. and {Prokhorov}, D. and {Prokoph}, H. and {P{"u}hlhofer}, G. and {Punch}, M. and {Quirrenbach}, A. and {Raab}, S. and {Rauth}, R. and {Reimer}, A. and {Reimer}, O. and {Renaud}, M. and {de los Reyes}, R. and {Rieger}, F. and {Rinchiuso}, L. and {Romoli}, C. and {Rowell}, G. and {Rudak}, B. and {Rulten}, C.~B. and {Safi-Harb}, S. and {Sahakian}, V. and {Saito}, S. and {Sanchez}, D.~A. and {Santangelo}, A. and {Sasaki}, M. and {Schandri}, M. and {Schlickeiser}, R. and {Sch{"u}ssler}, F. and {Schulz}, A. and {Schwanke}, U. and {Schwemmer}, S. and {Seglar-Arroyo}, M. and {Settimo}, M. and {Seyffert}, A.~S. and {Shafi}, N. and {Shilon}, I. and {Shiningayamwe}, K. and {Simoni}, R. and {Sol}, H. and {Spanier}, F. and {Spir-Jacob}, M. and {Stawarz}, {L}. and {Steenkamp}, R. and {Stegmann}, C. and {Steppa}, C. and {Sushch}, I. and {Takahashi}, T. and {Tavernet}, J. -P. and {Tavernier}, T. and {Taylor}, A.~M. and {Terrier}, R. and {Tibaldo}, L. and {Tiziani}, D. and {Tluczykont}, M. and {Trichard}, C. and {Tsirou}, M. and {Tsuji}, N. and {Tuffs}, R. and {Uchiyama}, Y. and {van der Walt}, D.~J. and {van Eldik}, C. and {van Rensburg}, C. and {van Soelen}, B. and {Vasileiadis}, G. and {Veh}, J. and {Venter}, C. and {Viana}, A. and {Vincent}, P. and {Vink}, J. and {Voisin}, F. and {V{"o}lk}, H.~J. and {Vuillaume}, T. and {Wadiasingh}, Z. and {Wagner}, S.~J. and {Wagner}, P. and {Wagner}, R.~M. and {White}, R. and {Wierzcholska}, A. and {Willmann}, P. and {W{"o}rnlein}, A. and {Wouters}, D. and {Yang}, R. and {Zaborov}, D. and {Zacharias}, M. and {Zanin}, R. and {Zdziarski}, A.~A. and {Zech}, A. and {Zefi}, F. and {Ziegler}, A. and {Zorn}, J. and {{.Z}ywucka}, N.},
	title = "{The H.E.S.S. Galactic plane survey}",
	journal = {Astron. Astrophys.},
	year = 2018,
	month = apr,
	volume = {612},
	eid = {A1},
	pages = {A1},
	doi = {10.1051/0004-6361/201732098},
	archivePrefix = {arXiv},
	primaryClass = {astro-ph.HE},
	adsurl = {https://ui.adsabs.harvard.edu/abs/2018A&A...612A...1H}
}

@ARTICLE{Malkov_2013,
       author = {{Malkov}, M.~A. and {Diamond}, P.~H. and {Sagdeev}, R.~Z. and {Aharonian}, F.~A. and {Moskalenko}, I.~V.},
        title = "{Analytic Solution for Self-regulated Collective Escape of Cosmic Rays from Their Acceleration Sites}",
      journal = {Astrophys. J.},
         year = 2013,
        month = may,
       volume = {768},
       number = {1},
          eid = {73},
        pages = {73},
          doi = {10.1088/0004-637X/768/1/73},
archivePrefix = {arXiv},
 primaryClass = {astro-ph.HE},
       adsurl = {https://ui.adsabs.harvard.edu/abs/2013ApJ...768...73M}
}

\end{document}